\documentclass{IEEEtran}
\usepackage{graphicx} 
\usepackage{enumitem}
\usepackage[colorlinks=true,linkcolor=blue, citecolor=blue, urlcolor=blue]{hyperref}
\usepackage{soul}
\usepackage{algorithm2e}
\usepackage{algorithmic}
\usepackage{xcolor}
\usepackage[space]{cite}
\usepackage[acronym]{glossaries}

\usepackage{tabularx}
\usepackage{booktabs}
\usepackage{ragged2e} 
\usepackage{makecell}
\usepackage{tikz}
\usepackage{multirow}
\usetikzlibrary{positioning, shapes, arrows.meta}
\usepackage{pifont}           
\usepackage{amsmath}
\usepackage{amssymb}
\usepackage{amsthm}
\usepackage[font=footnotesize,labelfont=bf]{subcaption}
\usepackage[font=small,skip=3pt]{caption}
\usepackage[compact]{titlesec}
\usepackage[T1]{fontenc}

\title{Rethinking Telemetry Design for Fine-Grained Anomaly Detection in 5G User Planes}

\author{
Niloy Saha\IEEEauthorrefmark{1},
Noura Limam\IEEEauthorrefmark{1},
Yang Xiao\IEEEauthorrefmark{1},
and Raouf Boutaba\IEEEauthorrefmark{1}\\
\{n6saha, noura.limam, yang.xiao, rboutaba\}@uwaterloo.ca, \IEEEauthorrefmark{1}University of Waterloo, Canada 
}

\newcommand{\kestrel}{\emph{Kestrel}\xspace}
\newcommand{\counter}{\emph{3GPP-PM}\xspace}
\newcommand{\dsamp}{\emph{$\Delta$-SMP}\xspace}

\setkeys{glslink}{hyper=false} 

\newcolumntype{Y}{>{\centering\arraybackslash}X}
\newcolumntype{R}{>{\raggedright\arraybackslash}X}

\newcommand{\signpost}[1]{\noindent \textbf{#1.}}
\newcommand{\indsignpost}[1]{\textbf{#1.}}
\newcommand{\itsignpost}[1]{\textit{#1.}}

\newacronym{SLO}{SLO}{Service Level Objective}
\newacronym{SLA}{SLA}{Service Level Agreement}
\newacronym{VPN}{VPN}{Virtual Private Network}
\newacronym{QoS}{QoS}{Quality of Service}
\newacronym{INT}{INT}{In-band Network Telemetry}
\newacronym{URLLC}{URLLC}{Ultra Reliable Low Latency Communications}
\newacronym{eMBB}{eMBB}{Enhanced Mobile Broadband}
\newacronym{mMTC}{mMTC}{Massive Machine Type Communications}
\newacronym{NF}{NF}{Network Function}
\newacronym{SDP}{SDP}{Service Demarcation Point}
\newacronym{3GPP}{3GPP}{3rd Generation Partnership Project}
\newacronym{RAN}{RAN}{Radio Access Network}
\newacronym{KPI}{KPI}{Key Performance Indicator}
\newacronym{KPM}{KPM}{Key Performance Metric}
\newacronym{E2E}{e2e}{end-to-end}
\newacronym{E2H}{E2H}{End-to-Hop}
\newacronym{IETF}{IETF}{Internet Engineering Task Force}
\newacronym{GTP}{GTP}{GPRS Tunnelling Protocol}
\newacronym{teid}{TEID}{Tunnel Endpoint Identifier}
\newacronym{SRv6}{SRv6}{Segment Routing IPv6}
\newacronym{ILP}{ILP}{Integer Linear Program}
\newacronym{PSTO}{PSTO}{Per-slice Telemetry Optimization}
\newacronym{MNO}{MNO}{Mobile Network Operator}
\newacronym{MPLS}{MPLS}{Multiprotocol Label Switching}
\newacronym{CU}{CU}{Centralized Unit}
\newacronym{UPF}{UPF}{User Plane Function}
\newacronym{snssai}{SNSSAI}{Single Network Slice Assistance Information}
\newacronym{qfi}{QFI}{QoS Flow Identifier}

\begin{document}

\maketitle

\begin{abstract}

Detecting QoS anomalies in 5G user planes requires fine-grained per-flow visibility, but existing telemetry approaches face a fundamental trade-off. Coarse per-class counters are lightweight but mask transient and per-flow anomalies, while per-packet telemetry postcards provide full visibility at prohibitive cost that grows linearly with line rate. Selective postcard schemes reduce overhead but miss anomalies that fall below configured thresholds or occur during brief intervals.

We present \kestrel, a sketch-based telemetry system for 5G user planes that provides fine-grained visibility into key metric distributions such as latency tails and inter-arrival times at a fraction of the cost of per-packet postcards. \kestrel extends Count-Min Sketch with histogram-augmented buckets and per-queue partitioning, which compress per-packet measurements into compact summaries while preserving anomaly-relevant signals. We develop formal detectability guarantees that account for sketch collisions, yielding principled sizing rules and binning strategies that maximize anomaly separability. Our evaluations on a 5G testbed with Intel Tofino switches show that \kestrel achieves $10\%$ better detection accuracy than existing selective postcard schemes while reducing export bandwidth by $10\times$.

\end{abstract}

\begin{IEEEkeywords}
5G, QoS, Network Slicing, Network Telemetry
\end{IEEEkeywords}

\section{Introduction} \label{sec:intro}

Modern 5G and beyond networks leverage \textit{network slicing} to provide dedicated logical networks for diverse applications, each with strict SLA requirements~\cite{ngmn-slicing16,3gpp_23.501,3gpp_22.261}. Data-plane components such as the User Plane Function (UPF) and O-RAN Centralized Unit (CU) aggregate traffic from multiple such slices onto hardware-accelerated platforms (e.g., P4 switches, SmartNICs, FPGAs). This makes them critical points for detecting QoS degradations that threaten slice SLAs. 


\indsignpost{The monitoring dilemma} 
Detecting QoS degradations in these data planes requires fine-grained, real-time telemetry. However, because thousands of flows share limited queues and on-chip memory, these systems typically only expose coarse per-class counters~\cite{3gpp_28.552}, or selective reports based on sampling or threshold triggers~\cite{sdfabric-int}, which obscure transient or per-flow anomalies. Programmable hardware can, in principle, support full per-packet visibility through \emph{postcard telemetry} that records queue depth, latency, and QoS identifiers for each packet. However, postcard export is prohibitively expensive: even at a modest 10 Gbps with 200 B packets, line-rate export would generate over 3 Gbps of telemetry (>30\% overhead), and production systems operate at hundreds of Gbps.

To make this challenge concrete, we focus on the UPF -- a critical 5G data plane function that forwards user traffic over GTP-U tunnels identified by \textit{tunnel endpoint identifiers (TEIDs)}. Each tunnel may carry multiple \textit{QoS flows (QFIs)} mapped to standardized service classes and enforced by a small set of shared hardware queues. When multiple TEIDs share a queue, standard 3GPP per-QFI counters~\cite{3gpp_28.552,open5gs} collapse fine-grained behavior into QFI-level averages. For example, counters polled every $1$s may report a healthy $2$ ms average delay at QFI-level, yet per-TEID histograms can reveal a brief spike in TEID A, with 6\% of packets delayed $>$20 ms. Postcards would expose this, but at unsustainable export cost.

This illustrates the fundamental challenge: \textit{how to obtain fine-grained (e.g., TEID-level) visibility under strict memory and bandwidth constraints?}


\indsignpost{Our approach} 
Our key insight is that data plane anomalies at the UPF manifest as distributional signatures (\S\ref{subsec:anomalies}): latency tails, sub-second spikes, per-TEID throughput violations, and oscillatory inter-arrival patterns. However, maintaining explicit per-TEID histograms is infeasible: each histogram requires multiple bins of state, which can quickly exhaust the limited memory available on hardware-accelerated data planes.

We present \kestrel, a telemetry system that provides per-TEID distributional visibility through a principled extension of Count-Min Sketch tailored to UPF anomaly detection. Unlike prior sketch-based systems~\cite{elasticsketch, nitrosketch} that focus on volume metrics (packet counts, heavy hitters), \kestrel augments sketch buckets to capture compact histograms of latency and inter-arrival times -- the distributional signals that distinguish different anomalies. To make distribution tracking robust, we develop a formal detectability framework that quantifies how sketch collisions and normal-traffic variation affect anomaly visibility. 
This analysis yields three practical design principles:
(1) per-QID sketch partitioning to limit cross-class interference,  
(2) adaptive binning that preserves sensitivity to rare events under normal load, and (3) principled sizing rules derived from anomaly detection requirements.


Our analysis shows that modest sketch dimensions ($w{=}512$, $d{=}3$) are sufficient to effectively detect diverse anomalies, including transient events as short as $50$ ms. \kestrel{} exports compact summaries (${\approx}0.15\%$ of user-plane traffic), achieving over \textit{$10\times$ lower telemetry volume} than existing selective-postcard schemes~\cite{sheng2021deltaint} while improving anomaly detection accuracy by \textit{$10\%$} on mixed workloads.


\indsignpost{Contributions} We make the following contributions:
\begin{itemize}[leftmargin=*]
\item We characterize the telemetry requirements for UPF anomaly detection, showing that representative anomalies require per-TEID distributional visibility such as latency tails, inter-arrival patterns, and sub-second resolution (\S\ref{sec:background}).

\item We design a histogram-augmented sketch data structure that tracks per-TEID distributions using shared buckets rather than per-flow state, enabling operation within the memory constraints of programmable switches (\S\ref{sec:design}).

\item We develop formal detectability guarantees for sketch-based anomaly detection, and derive principled sizing rules for sketch dimensions and bin placement that ensure weak anomalies remain visible despite measurement noise (\S\ref{sec:analysis}).

\item We implement \kestrel on Intel Tofino switches, and demonstrate that it achieves $10\%$ better anomaly detection accuracy than selective postcard sampling while reducing export bandwidth by $10\times$ (\S\ref{sec:evaluation}). 
\end{itemize}

\section{Characterizing the Fine-Grained Telemetry Problem in 5G Data Planes} \label{sec:background}


Modern 5G networks increasingly implement key data plane functions, including the User Plane Function (UPF) and O-RAN Centralized Unit (CU), on hardware-accelerated platforms~\cite{hybridp4pipelines-tmc23,blink-noms24,p4ufp-sosr21,p4upf-atc24}. These functions aggregate traffic from multiple network slices, which makes them critical points for detecting QoS degradations that threaten SLA compliance. We define any data plane event that threatens SLA compliance as an \textit{anomaly}, whether it stems from malicious activity, misconfiguration, or disruptive traffic patterns.

The central challenge in detecting such anomalies is a fundamental \emph{visibility gap}: these systems typically expose only coarse-grained counters or heavily sampled telemetry, which fail to capture the fine-grained, transient signals required for reliable detection. In UPFs, this arises from aggregation of per-session statistics; in CUs, it manifests as coarse event reports that obscure per-UE dynamics~\cite{tcran-jsac24}.

We focus on the UPF as a representative case, developing and evaluating our approach on a UPF testbed. However, the visibility gap and resulting telemetry requirements apply broadly across 5G data plane components. We first characterize the architectural and resource constraints that create this gap (\S\ref{subsec:visibility_gap}), then demonstrate empirically what telemetry granularity is required for reliable anomaly detection (\S\ref{subsec:anomalies}).

\subsection{The Visibility Gap in UPF Telemetry} \label{subsec:visibility_gap}

The visibility gap stems from a fundamental mismatch between QoS enforcement requirements and hardware constraints. UPFs forward traffic using GTP-U tunnels identified by TEIDs, with service classes distinguished by QFI and enforced by QERs. To sustain line-rate throughput, UPFs are increasingly offloaded to programmable hardware (e.g., P4 switches, SmartNICs) that provide typically 8-100 egress queues per port~\cite{hybridp4pipelines-tmc23,p4ufp-sosr21,p4upf-atc24}. Thousands of active TEIDs must therefore multiplex onto these shared queues, creating QoS interdependencies: a single aggressive TEID can degrade performance for others sharing its queue, making per-TEID monitoring essential for anomaly attribution.

However, tracking per-TEID metrics explicitly is infeasible: programmable data planes provide only a few MB of SRAM, insufficient for maintaining counters across thousands of concurrent TEIDs. Existing telemetry systems work around this memory constraint:

\indsignpost{Aggregate counters} Standard 3GPP performance metrics~\cite{3gpp_28.552}, implemented in systems such as Open5GS~\cite{open5gs}, report per-QFI packet/byte counts, loss, and average delay every few seconds. By aggregating across TEIDs, they reduce state requirements but collapse per-packet dynamics into coarse aggregates that mask TEID-level variation.

\indsignpost{Postcard telemetry} Programmable data planes can generate stateless per-packet metadata~\cite{handigol2014,sdfabric-int,intel-int20} containing queue depth, timestamps, and identifiers. Postcards avoid maintaining per-TEID state, but impose prohibitive export bandwidth: at 10 Gbps with 200B packets, line-rate export exceeds 3 Gbps (>30\% overhead). Recent systems adopt change-triggered sampling ($\Delta$-SMP)~\cite{sheng2021deltaint,lint-im21}, exporting telemetry only when change in metrics exceed a threshold $\Delta$, but events which remain below thresholds or overlap with others are missed.

Neither approach provides what anomaly detection requires: \textit{temporal granularity} to capture transient events, \textit{spatial granularity} to isolate per-TEID behavior, and \textit{distributional visibility} into queue dynamics beyond simple averages. Other telemetry systems face similar limitations (\S\ref{sec:related_work}). We demonstrate these requirements empirically next.

\subsection{Empirical Characterization of Anomaly Signatures} \label{subsec:anomalies}

\begin{figure*}[!ht]
    \centering
    \setlength{\tabcolsep}{1pt} 
    \begin{tabular}{cc}
    \begin{subfigure}[t]{0.49\textwidth}
        \includegraphics[width=\linewidth]{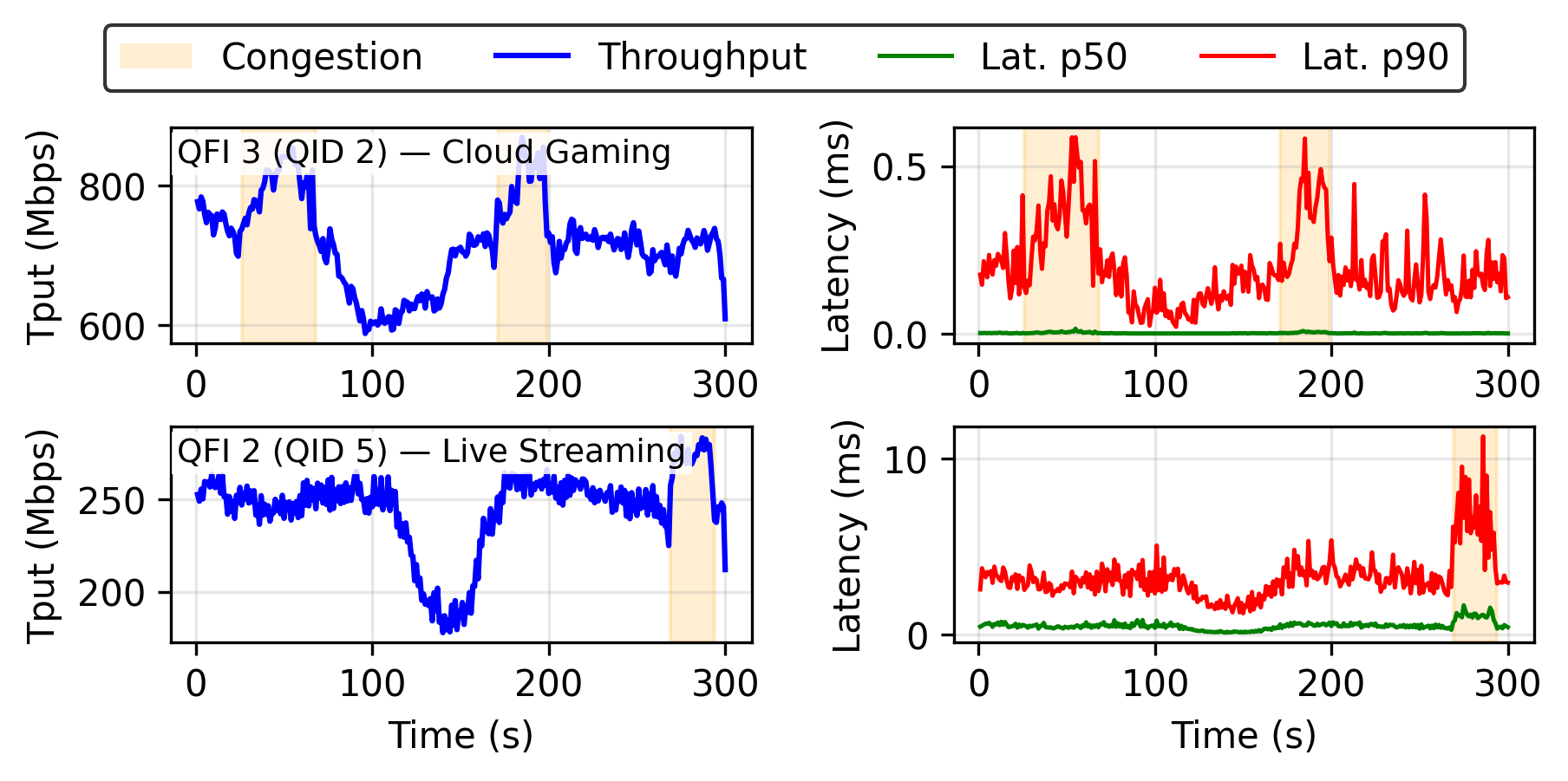}
        \caption{Congestion}
        \label{fig:congestion}
    \end{subfigure} &
    \begin{subfigure}[t]{0.49\textwidth}
        \includegraphics[width=\linewidth]{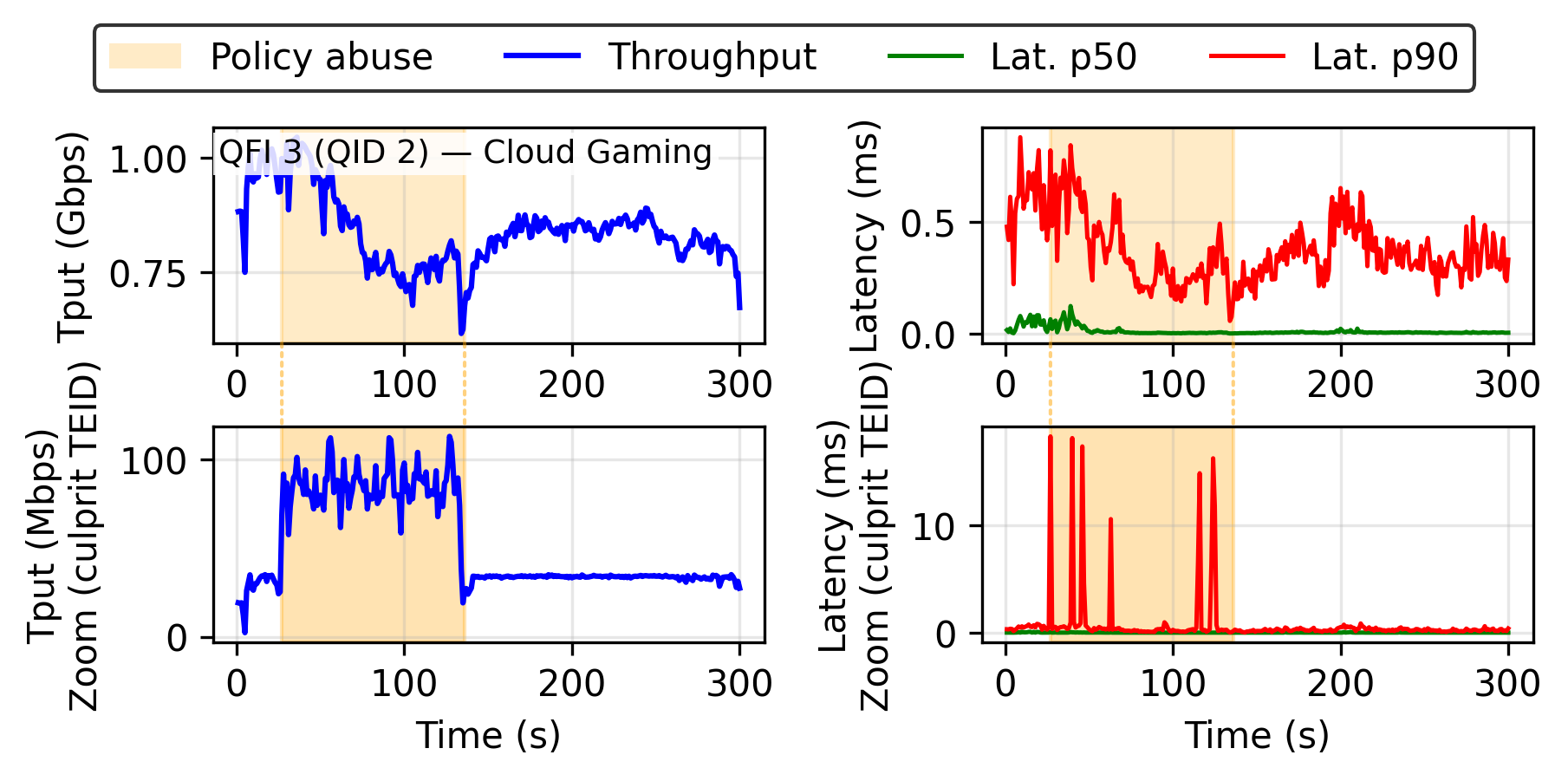}
        \caption{Policy abuse}
        \label{fig:policy}
    \end{subfigure} \\
    
    \begin{subfigure}[t]{0.49\textwidth}
        \includegraphics[width=\linewidth]{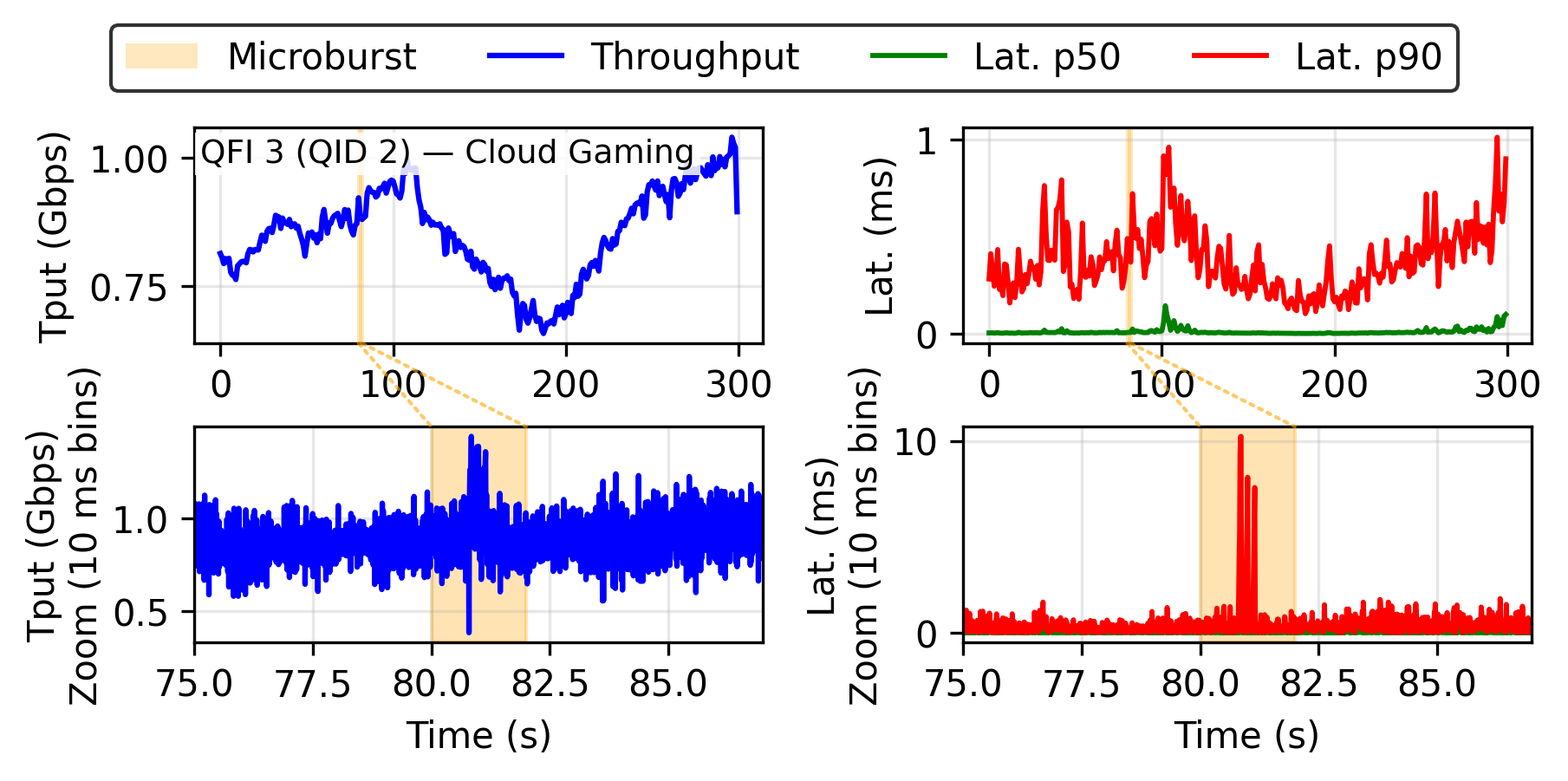}
        \caption{Microburst}
        \label{fig:microburst}
    \end{subfigure} &
    \begin{subfigure}[t]{0.49\textwidth}
        \includegraphics[width=\linewidth]{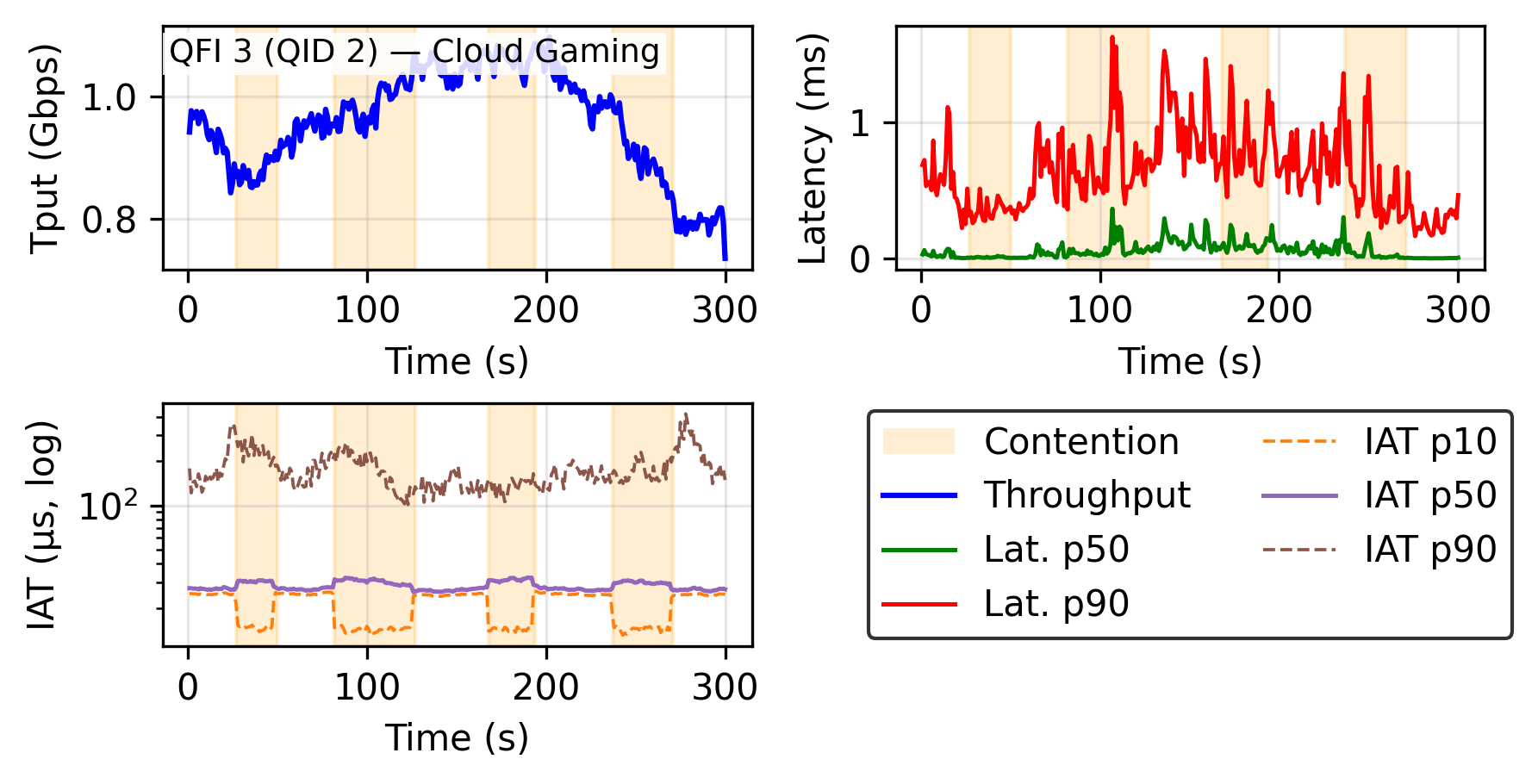}
        \caption{Contention}
        \label{fig:contention}
    \end{subfigure}
    \end{tabular}
    
    \caption{Anomaly signatures require appropriate monitoring granularity.  
    (a) Congestion produces sustained tail latency on specific QFIs.  
    (b) Policy abuse appears benign in QFI aggregates but is visible at the culprit TEID.  
    (c) Microbursts vanish in second-level averages yet emerge with sub-second resolution.  
    (d) Contention induces distributed oscillations in inter-arrival times across flows.}
    \label{fig:anomalies}
\end{figure*}

To validate these requirements and understand what telemetry signals enable reliable anomaly detection, we inject controlled anomalies into our Tofino-based UPF testbed carrying a mix of application traffic: cloud gaming, live streaming, video conferencing, and IoT. Each application maps to standardized QFIs, with multiple TEIDs per application sharing queues. We implement eight queues\footnote{Our Tofino-1 testbed operates at 10~Gbps with 8 egress queues per port.} with strict-priority and weighted round-robin scheduling, plus TrTCM-based policing~\cite{p4ufp-sosr21,p4upf-atc24}.

Figure~\ref{fig:anomalies} shows four representative anomaly types and their detection requirements. \emph{Congestion} from sustained surges produces long-tail latencies concentrated on specific QFIs. (Fig.~\ref{fig:congestion}). \emph{Policy abuse}, where TEIDs exceed QER allocations (e.g., by remapping to higher-priority classes), appears benign at the QFI level but reveals throughput violations and latency spikes when zoomed in to the TEID level (Fig.~\ref{fig:policy}).  \emph{Microbursts} from bursty video streams manifest as millisecond-scale spikes that vanish in one-second QFI aggregates but emerge at finer temporal resolution (Fig.~\ref{fig:microburst}). \emph{Contention} from competing backhaul traffic induces oscillatory inter-arrival patterns distributed across flows (Fig.~\ref{fig:contention}).  

These anomalies share a critical property: \textit{their signatures become visible only with appropriate measurement granularity}. Specifically, effective detection requires: 
\begin{enumerate}[leftmargin=*]
\item \emph{Sub-second temporal resolution} to capture transient bursts.
\item \emph{Per-TEID spatial resolution} to attribute misbehavior within shared queues.
\item \emph{Distributional features} such as latency tails and inter-arrival patterns rather than simple averages.
\end{enumerate} 

The challenge is designing a telemetry primitive that preserves these signals while remaining lightweight enough for carrier-scale deployment.

\section{Design of \kestrel} \label{sec:design}

\subsection{Design Objectives} \label{subsec:goals}

Section~\ref{sec:background} established that reliable anomaly detection requires per-TEID distributions at sub-second resolution without the memory overhead of explicit tracking or the bandwidth cost of continuous export. Addressing this challenge requires meeting three objectives simultaneously:

\begin{enumerate}[leftmargin=*]
\item \textbf{Bounded export cost.}  
Telemetry overhead must remain fixed even as traffic rates and active TEIDs grow, eliminating the bandwidth explosion that makes postcard export untenable at carrier scale.

\item \textbf{Per-TEID distributional fidelity.}  
Telemetry must retain anomaly signatures such as latency tails, inter-arrival irregularities, and color counts, for each TEID, rather than collapsing behavior into slice-level averages (\S\ref{subsec:anomalies}).

\item \textbf{Pipeline-constrained efficiency.}  
The telemetry primitive must fit comfortably within switch memory and execute within a single pipeline pass, avoiding recirculation, external memory, or coordination across stages. This ensures deployability on commodity programmable hardware such as SmartNICs and Tofino switches, while leaving room for other 5G data plane functions.
\end{enumerate}

\subsection{Design Challenges and Solutions} \label{subsec:challenges}

Meeting the objectives in \S\ref{subsec:goals} is not straightforward. One natural alternative is to optimize selective postcard sampling, for example, with adaptive thresholds or smarter triggers. Such schemes fundamentally violate Objective~(1): their export cost remains traffic dependent, flooding collectors precisely during bursts when visibility is most critical. We instead pursue sketch-based summaries that bound export volume regardless of conditions. For this, we identify two main challenges:

\indsignpost{Challenge 1: From volumes to distributions} State-of-the-art sketches such as ElasticSketch~\cite{elasticsketch} and NitroSketch~\cite{nitrosketch} excel at estimating flow volumes and identifying heavy hitters. They achieve high accuracy at scale, but remain tied to \emph{volume metrics}: packet counts, byte counts, and flow sizes. What they cannot capture is how packets are distributed across latency or inter-arrival bins, the very signals that distinguish anomalies such as microbursts, contention, or policy abuse.  

\kestrel addresses this gap by \emph{extending Count–Min Sketch (CMS)} to preserve distribution signals. CMS is an attractive foundation because it offers (i) clean analytical error bounds, (ii) a simple hash-and-increment update model that maps directly to ASIC pipelines, and (iii) a lightweight footprint that leaves memory and ALU budget for essential UPF functions such as policing, queuing, and scheduling. While CMS does not optimize overestimation bias, its simplicity makes it the right starting point.  

To further mitigate bias, \kestrel maintains \emph{per-QID sketches} rather than a single global structure (\S\ref{sec:architecture}). Since QIDs group flows with similar QoS targets, partitioning reduces cross-class collisions and makes error bounds less pessimistic in practice.  Each CMS bucket is \emph{augmented} to hold not only packet and byte counters but also compact histograms of latency, inter-arrival times, and policing colors. These enriched buckets yield one-second summaries that preserve the distribution signals required for anomaly detection. The challenge lies not in augmenting the sketch buckets but in \emph{configuring them correctly}: bin edges must cap diagnostic occupancy, and sketch width $w$ and depth $d$ must be sized to ensure anomalies remain visible despite collisions. We formalize these requirements in \S\ref{sec:analysis}, deriving detectability guarantees and sizing rules that make histogram-augmented sketches practical.

\indsignpost{Challenge 2: Capturing timing without per-flow state}  
Objective~(2) also requires preserving inter-arrival time (IAT) patterns, which are crucial for distinguishing contention from congestion. A naïve design would timestamp each active flow, requiring $O(F)$ state for $F$ TEIDs, far beyond ASIC budgets and violating Objective~(3).  
\kestrel instead timestamps \emph{buckets}, not flows. Each CMS bucket stores the arrival time of its last packet. When the next packet (possibly from a colliding flow) arrives, the IAT is computed relative to this stored timestamp and recorded into the bucket's histogram. This reduces state to $O(d\cdot w)$ rather than $O(F)$. While collisions introduce some noise, our analysis (\S\ref{sec:analysis}) shows that the diagnostic bins remain sparse under baseline traffic, so anomalies still stand out reliably. 

Therefore, adapting sketches for UPF anomaly detection requires rethinking both the \emph{signal model} (distribution vs.\ volume) and the \emph{state model} (bucket vs.\ per-flow). We next present \kestrel{}'s data structure that realizes this design.

\subsection{Data Structure and Operations} \label{sec:architecture}

\begin{figure*}[!ht]
    \centering
    \includegraphics[width=0.9\linewidth]{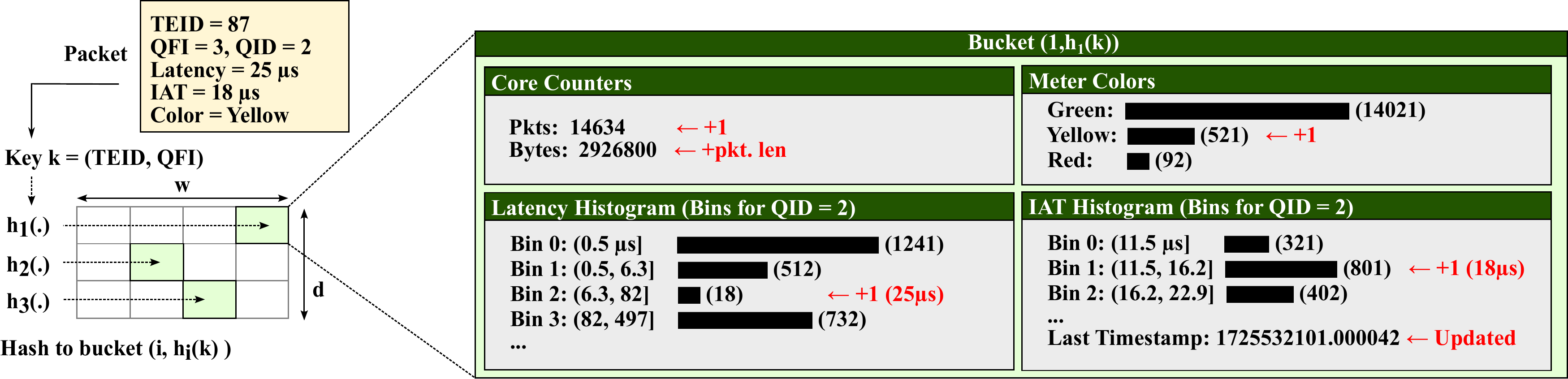}
    \caption{Data structure and operations of \kestrel. Buckets extend CMS to record packet/byte totals, latency histograms, IAT histograms, and policer colors. Example shows an arriving packet (TEID=87, QFI=3, QID=2, latency=25 µs, IAT=18 µs, color=yellow) being hashed to a bucket: totals and color counters are incremented, latency/IAT bins are updated using QID-specific edges, and the last is timestamp refreshed (updates in red).}
    \label{fig:sketch}
\end{figure*}

\indsignpost{Architecture overview} \kestrel combines two design choices that make sketching effective in the UPF setting. 

First, \textit{per-QID partitioning}: rather than one global sketch, \kestrel maintains a separate sketch per QID (8 in our prototype). Packets mapped to the same queue by UPF scheduling policies typically share QoS targets (e.g., low-latency cloud gaming vs. best-effort bulk). Per-QID partitioning reduces cross-class collisions, and enables queue-specific binning (a 1\,ms latency threshold is meaningful for gaming but irrelevant for best-effort). While this increases the number of sketches, each sketch can be kept small: as we show in \S\ref{sec:analysis} and \S\ref{sec:implementation}, sketches with only $w{=}512$ and $d{=}3$ suffice for robust anomaly detection. This makes per-QID partitioning practical, since the total memory cost remains well within hardware budgets.

Second, \textit{CMS with enriched buckets} to capture distributional signals. \kestrel maintains $d$ rows and $w$ buckets per row, with independent hash functions $h_1,\dots,h_d$ over QoS flow keys $k=(\text{TEID},\text{QFI})$. As shown in Figure~\ref{fig:sketch}, each bucket $j_i=h_i(k)$ stores four components: (a) \emph{core counters} for packets and bytes, (b) \emph{meter colors} tallying TrTCM outcomes (green, yellow, red), (c) a \emph{latency histogram} updated from per-packet sojourn time, and (d) an \emph{IAT histogram} updated from the bucket's timestamp register. These enriched buckets allow \kestrel to summarize traffic distributions compactly.

\indsignpost{Packet updates} On each arriving packet with key \( k=(\text{TEID},\text{QFI}) \) and latency \( t_s \), \kestrel computes the bucket indices \( j_i=h_i(k) \) for each row \( i \). For each corresponding bucket \( (i, j_i) \), it then: 1) increments the packet and byte counters, 2) maps the latency \( t_s \) to a bin using QID-specific edges, 3) computes the Inter-Arrival Time (IAT) as the difference from the bucket's last-seen timestamp, updates the corresponding IAT bin, and overwrites the timestamp, and 4) increments the color counter corresponding to the packet's TrTCM marking.

\indsignpost{Export and query model} At the end of each 1\,s window, the control plane collects one compact record per bucket. To estimate the feature vector for a specific flow with key $k$, it queries all $d$ buckets $h_i(k)$ to which the flow was mapped. The estimate for each feature (e.g., the count in latency bin \#3) is the minimum across the $d$ candidate buckets, following CMS query semantics. The control plane, which already maintains TEID--QFI mappings for bearer/session management, uses these mappings to reconstruct per-flow distributions suitable for ML-based anomaly detection. This model keeps the switch pipeline lightweight while ensuring anomaly-relevant signals are exposed at second-level cadence.

\indsignpost{Illustrative Example (Figure~\ref{fig:sketch})} Consider a GTP-U packet arriving at the UPF with TEID=87 and QFI=2, experiencing a sojourn time of $25\,\mu$s and an inter-arrival gap of $18\,\mu$s. The UPF maps this flow to QID=3 according to its scheduling policies. \kestrel processes this packet as follows: First, it forms the sketch key $k$ from TEID+QFI and applies $d$ hash functions to map $k$ to buckets $(i,j_i)$ across all rows. Second, it retrieves the QID=3-specific bin edges; for latency ($\{0.5, 6.3, 82, \ldots, 4970\}\,\mu$s) and IAT ($\{11.5, 16.2, 22.9, \ldots, 2.7{\times}10^6\}\,\mu$s). Finally, for each bucket $(i,j_i)$, it: increments packet/byte counters; assigns the $25\,\mu$s latency to bin~2 (as shown in Fig.~\ref{fig:sketch}); computes and records the $18\,\mu$s IAT (updating bin~2 and refreshing the timestamp); and updates the color counter based on meters.

\section{Parameterizing Kestrel} \label{sec:analysis}

\kestrel{}'s design raises three practical questions:  
(i) how to ensure detectability when anomalies only occupy a small region of the distribution,  
(ii) how to size the sketch parameters to meet detection targets, and  
(iii) how to place bin boundaries and handle distribution drift.  
We address each in turn.

\subsection{Detectability Guarantees}

We first establish when anomalies become visible.
As discussed in \S\ref{sec:background}, anomalies typically concentrate in certain regions of the latency or inter-arrival distributions.
We designate these regions (the latency tail and IAT head bins) as the \emph{diagnostic region}~$T$. Let $N_T$ denote the number of packets in $T$ under normal traffic. Because $T$ is kept sparse (\S\ref{subsec:binning}), any anomaly-induced increase~$\Delta_T$ in that region becomes readily detectable.

However, two factors can obscure this signal. First, only a fraction of anomaly packets may actually fall in $T$; we denote the spillover fraction by~$\beta$ ($\beta{=}0$ means all land in~$T$).
Second, sketch collisions introduce noise on the order of~$\varepsilon N_T$, where $\varepsilon{\approx}e/w$ is the CMS error rate.
The following theorem establishes detectability while accounting for both effects:

\indsignpost{Detectability Condition}
For a flow~$k$, an anomaly that adds~$\Delta_T$ packets to its diagnostic bins~$T$ 
and~$\beta\Delta_T$ outside~$T$ is detectable with probability at least~$1-\delta$ when
\[
\Delta_T \;>\; 
\frac{\varepsilon N_T x_k + (x_{k,T}+\varepsilon N_T)\,\varepsilon N'}
     {x_{k,\bar T} - \varepsilon N_T - \beta(x_{k,T}+\varepsilon N_T)} ,
\]
where $x_k$ is the total mass of~$k$ in the window, 
$x_{k,T}$ and $x_{k,\bar T}$ are its baseline masses in and outside~$T$, 
$N'$ is the total packet mass, and~$\varepsilon{\approx}e/w$ is the CMS error rate. (Proof in Appendix~A.)

\itsignpost{Interpretation} Detection succeeds when the anomaly lift~$\Delta_T$ within~$T$ exceeds the collision floor~$\varepsilon N_T$
after accounting for spillover~$\beta$.
For example, with~$N{=}10^6$ pkt/s and diagnostic cap~$\rho{=}1\%$ ($N_T{=}10^4$),
$w{=}512$ gives~$\varepsilon N_T{\approx}53$ pkts.
At~$\beta{=}0.3$, the required~$\Delta_T{\gtrsim}76$ pkts.
Even a small $50$ms microburst at $50$kpps injects ${\approx}2500$ pkts, of which $70\%$ ($\Delta_T{\approx}1,750$) fall in~$T$, well above the threshold.

\subsection{Choosing Width and Depth} \label{sec:sketch_parameters}

The detectability condition translates directly into concrete sizing rules for the sketch parameters: 
width~$w$ and depth~$d$.

\indsignpost{Width Requirement}
To detect all anomalies that inject at least~$\Delta_T^{\min}$ packets into the diagnostic region~$T$, 
the sketch width must satisfy

$$
w \;\ge\; 
\frac{e\,N_T^{\max}}{(1-\beta_{\max})\,\Delta_T^{\min}},
$$
where~$N_T^{\max}$ is the maximum diagnostic occupancy (set by binning) and~$\beta_{\max}$ bounds the spillover fraction. For instance, with~$N_T^{\max}{=}10^4$ and~$\beta_{\max}{=}0.3$, detecting anomalies that inject ${\sim}80$ packets into~$T$ 
requires~$w{\ge}485$; we use~$w{=}512$ in practice, a configuration empirically validated in~\S\ref{subsec:cost_effectiveness}.

\indsignpost{Depth Requirement}
Let~$K{=}|T|$ be the number of diagnostic bins.  
If
$$
d \;\ge\; \left\lceil \ln\!\frac{K+1}{\zeta}\right\rceil,
$$
then, with probability at least~$1-\zeta$, all CMS bounds needed for detection hold simultaneously within a window.
In practice,~$K$ is small (typically 2–3 bins for latency tails and IAT heads). 
With~$d{=}3$, these bounds hold with~$>99\%$ probability per window, 
and since most anomalies persist across multiple windows, the overall miss probability becomes negligible.

\subsection{Binning and Drift} \label{subsec:binning}

Detectability depends not only on sketch width~$w$ but also on the background occupancy~$N_T^{\max}$ 
of the diagnostic region~$T$.  
Because~$N_T^{\max}$ is set by bin placement, careful binning is crucial: 
if too much baseline traffic falls in~$T$, the collision noise~$\varepsilon N_T$ grows, 
requiring larger widths to maintain the same detection guarantee.

\indsignpost{Target-Occupancy Binning}  
\kestrel{} controls~$N_T^{\max}$ using \emph{target-occupancy binning}: 
bin edges are placed to cap baseline occupancy in~$T$ at a target fraction~$\rho$ 
(typically~1–2\,\% of total packets).  
Setting the latency-tail boundary at the $(1{-}\rho)$-quantile and the IAT-head boundary at the~$\rho$-quantile 
ensures~$N_T^{\max}{=}\rho N$ (see Appendix for derivation).  
This design maximizes anomaly separability by keeping diagnostic bins nearly empty under normal conditions. We show that this
strategy outperforms alternative binning schemes in~\S\ref{subsec:microbenchmarks}.

\indsignpost{Handling Drift}  
Traffic distributions naturally drift over time, potentially increasing~$N_T$ beyond~$\rho N$.  
Left uncorrected, this raises the required width roughly in proportion to occupancy.  
To preserve sensitivity, the control plane monitors~$N_T$ from exported sketches
and triggers \emph{re-binning} when~$\hat N_T/N > \rho$, \emph{during anomaly-free periods}. Our P4 implementation supports this dynamically: bin edges are stored as runtime-programmable table entries keyed by bin ranges and queue ID, allowing updates via P4Runtime without pipeline disruption.

\indsignpost{Take-away} Our analysis yields three configuration rules:
(i) reserve a small diagnostic region (2–3 bins for latency tails and IAT heads); (ii) cap baseline occupancy using target-occupancy binning ($\rho{\le}2\%$) with drift adaptation; and (iii) set sketches to~$w{\approx}512$, $d{=}3$–4 under typical workloads.  
These parameters ensure~$\Delta_T \!\gg\! \varepsilon N_T$ for practical anomalies, providing robust detection at modest memory cost.  
With these settings, \kestrel{} exports compact, anomaly-revealing summaries light enough for deployment in programmable UPFs.

\section{Implementation} \label{sec:implementation}

\begin{table*}[!ht]
\centering
\begin{minipage}[t]{0.45\textwidth}
\centering
\setlength{\tabcolsep}{3.5pt}
\renewcommand{\arraystretch}{1.05}
\caption{Anomaly scenarios used in our evaluation: single anomalies (1–4) and mixed anomalies (5).}
\label{tab:anomalies}
\begin{tabular}{c l l c c}
  \toprule
  \textbf{Scenario} & \textbf{Anomaly} & \textbf{Target} & \textbf{Dur.} & \textbf{Freq.} \\
  \midrule
  1 & Microburst   & One/more TEIDs   & 0.3--1.0\,s    & 10--30\,s \\
  2 & Congestion   & One/more QFIs    & 25--40\,s      & 60--120\,s \\
  3 & Contention   & Many QFIs/TEIDs  & 12--30\,s      & 90--150\,s \\
  4 & Policy abuse & Selected TEIDs   & 30--60\,s      & 150--300\,s \\
  5 & Mixed        & Multiple         & As above       & As above \\
  \bottomrule
\end{tabular}
\end{minipage}
\hfill
\begin{minipage}[t]{0.52\textwidth}
\centering
\setlength{\tabcolsep}{3.5pt}
\renewcommand{\arraystretch}{1.05}
\caption{Telemetry approaches evaluated in our testbed, summarizing their switch operations, collection mechanisms, and asymptotic overhead.}
\label{tab:pipelines}
\begin{tabularx}{\columnwidth}{lXXX}
  \toprule
  \textbf{Baseline} & \textbf{Switch} & \textbf{Collector} & \textbf{Overhead} \\
  \midrule
  \textbf{3GPP-PM} 
  & Per-QFI counters
  & Poll every 1s 
  & \textit{Low}, $O(\#\text{QFIs})$ \\
  \addlinespace
  \textbf{Kestrel} 
  & Per-QID sketches
  & Query every 1s
  & \textit{Medium}, $O(wd)$ \\
  \addlinespace
  \textbf{$\Delta$-SMP} 
  & Postcard sampling 
  & Collect \& aggregate every 1s
  & \textit{High}, $O(\#\text{pkts})$ \\
  \bottomrule
\end{tabularx}
\end{minipage}
\end{table*}

\indsignpost{Testbed and Data plane} We prototype \kestrel on a UfiSpace S9180-32X switch equipped with an Intel Tofino ASIC. The switch implements UPF functions including queuing, scheduling, and TrTCM metering, and connects over 10\,GbE NICs to four Ubuntu~22.04 servers. These servers act as: (i) a GTP-U traffic generator that allows fine-grained control over TEID/QFI mappings, (ii) a control plane host that programs P4 tables and configures runtime telemetry, (iii) a collector for postcards and sketch registers, and (iv) an ML pipeline server for anomaly detection.  

The data plane implementation consists of $\sim$2K lines of P4 in the egress pipeline. The design supports all three telemetry modes: 3GPP-PM counters, postcards (via packet mirroring), and sketches. It fits within a single pipeline pass ($\sim$7 stages) without recirculation. Sketches use $d{=}3$ rows and $w{=}512$ buckets (\S\ref{sec:analysis}); each bucket stores counters, 8-bin latency and IAT histograms, and TrTCM color tallies. Per-QID bin tables are runtime-programmable, enabling dynamic reconfiguration. Queues implement strict-priority and WRR scheduling across QIDs to emulate realistic UPF service classes.

\indsignpost{Workloads and Anomalies} Traffic workloads combine a 5G operator trace~\cite{korean-dataset} (cloud gaming, live streaming, buffered streaming) with additional traffic types we collected for IoT, VoIP, and best-effort applications. The generator sustains 1--4\,Gbps aggregate load across up to 100~UEs $\times$ 9~QFIs ($\approx$900 flows)\footnote{Our prototype generator sustains a few Gbps; higher loads are feasible with a DPDK-based design, which we leave to future work.}. We inject four representative anomaly scenarios: \textit{microbursts} (short high-rate bursts), \textit{congestion} (sustained overload), \textit{contention} (shared backhaul bottlenecks), and \textit{policy abuse} (QFI remapping), as well as mixed cases. Table~\ref{tab:anomalies} summarizes their scope, durations, and injection frequencies. Each anomaly ultimately degrades QoS for affected UEs.

\indsignpost{Telemetry export and processing} Sketches are exported every $1$s via the Barefoot gRPC API, while postcards are captured by a dedicated collector with a memory-mapped ring buffer for high-throughput, zero-copy capture. All telemetry modes are aligned to one-second windows for fair comparison. Feature extraction proceeds according to the intrinsic visibility of each mode. Sketches yield per-(TEID,QFI) distributions by querying the $d$ rows for each key and applying CMS query semantics. Postcards provide per-(TEID,QFI) distributions directly from INT metadata (latency, IAT, color). Counters expose only QFI-level aggregates. From these, we construct feature vectors including traffic volume, latency percentiles and tail fractions, inter-arrival head fractions, policer color ratios, and contextual metrics such as TEID counts per QFI.

\indsignpost{Anomaly detection pipeline} We implement an anomaly detection pipeline incorporating four dedicated detectors that target congestion, contention, microbursts, and policy abuse. Each detector is implemented as an XGBoost model trained on baseline and anomaly-injected traces, enabling robust classification between normal and abnormal traffic patterns. As a result, each detector is allowed to focus on anomaly-specific behaviors. For instance, the microburst detector leverages sub-second temporal spikes, while the contention detector uses distributional shifts across flows. To ensure robustness, models are validated using temporally blocked cross-validation, preventing leakage between training and test windows.
\section{Evaluation} \label{sec:evaluation}

We evaluate \kestrel{} to answer four questions:
\begin{enumerate}[leftmargin=*]
\item How \textit{accurately} can it detect diverse anomalies? 
\item How \textit{cost-effective} is it compared to existing approaches?
\item How \textit{quickly} can it detect emerging anomalies?
\item Is it \textit{practical} for deployment on programmable switches?
\end{enumerate}

\indsignpost{Baselines} We compare \kestrel{} against two representative telemetry approaches:
(i) \counter, the standards-defined per-QFI performance measurements~\cite{3gpp_28.552,open5gs}, which expose per-QFI packet and byte counts, average delay, and loss; and
(ii) \dsamp, selective postcard sampling that exports telemetry only when monitored metrics change beyond a configured $\Delta$ threshold~\cite{sheng2021deltaint,lint-im21}, thereby reducing postcard volume relative to per-packet export. Table~\ref{tab:pipelines} presents the key characteristics of these baselines alongside \kestrel{}, highlighting their switch operations, collection mechanisms, and asymptotic overhead. These baselines capture the design spectrum -- from coarse-grained low-overhead (\counter) to fine-grained high-overhead (\dsamp).

We next evaluate these approaches along the four axes above:
accuracy (\S\ref{subsec:anomaly_detection}),
cost-effectiveness (\S\ref{subsec:cost_effectiveness}),
responsiveness (\S\ref{subsec:responsiveness}), and
practicality (\S\ref{subsec:microbenchmarks}).

\subsection{Anomaly Detection Accuracy} \label{subsec:anomaly_detection}

\begin{figure*}[!t]
  \centering

  \begin{minipage}[b]{0.6\linewidth}
    \centering
    \vspace{0pt}
    \begin{subfigure}[b]{0.6\linewidth}
      \centering
      \includegraphics[width=\linewidth]{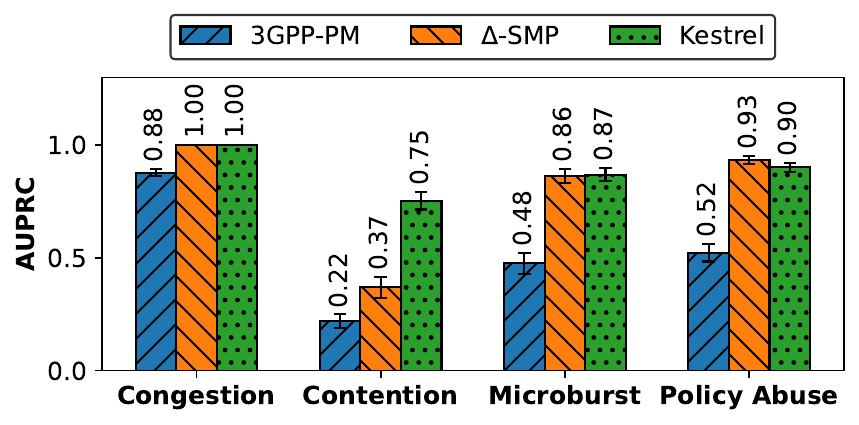}
      \caption{Single anomaly type (Scenarios~1--4).}
      \label{fig:auprc_per_head}
    \end{subfigure}%
    \hfill
    \begin{subfigure}[b]{0.38\linewidth}
      \centering
      \includegraphics[width=\linewidth]{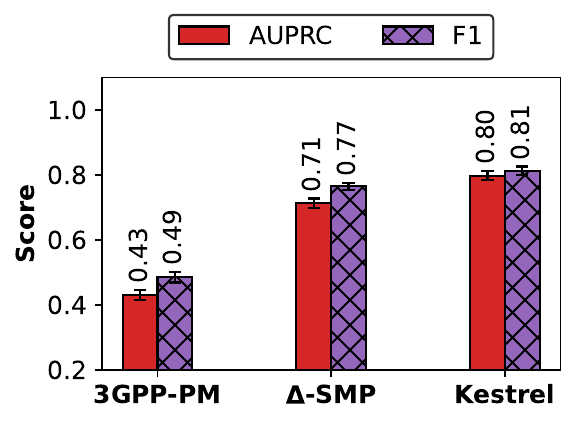}
      \caption{Mixed anomalies (Scenario~5).}
      \label{fig:auprc_mixed}
    \end{subfigure}

    \captionof{figure}{Detection accuracy across telemetry approaches. 
    (a) Single-anomaly scenarios (Scenarios~1–4): Continuous, distributional visibility (\kestrel{}) outperforms coarse counters and threshold-triggered sampling.
    (b) Mixed anomalies (Scenario~5): \kestrel{} sustains high AUPRC and F1, demonstrating robustness to concurrent anomaly types.}
    \label{fig:auprc_combined}
  \end{minipage}
  \hspace{0.02\linewidth}
  \begin{minipage}[b]{0.34\linewidth}
    \centering
    \vspace{0pt}
    \includegraphics[width=\linewidth]{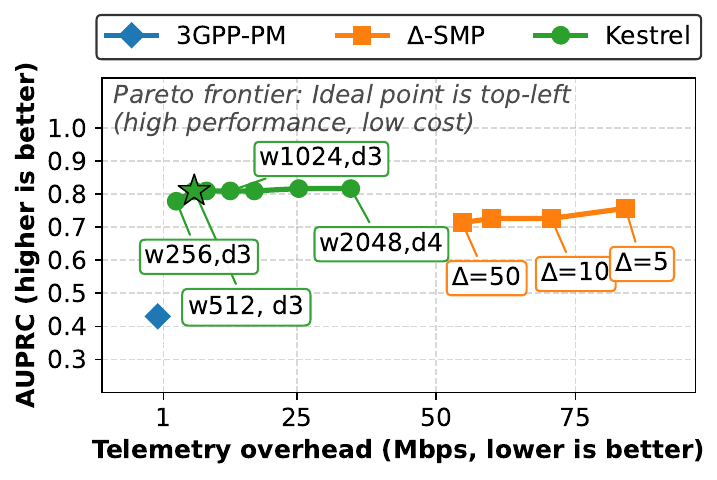}

    \captionof{figure}{Pareto of AUPRC vs telemetry cost.
    \kestrel{}'s best ($w{=}512,d{=}3$) reaches AUPRC~0.81 at $\sim$6\,Mbps, 
    \mbox{2$\times$} higher than \counter and \mbox{10$\times$} cheaper than \dsamp.}
    \label{fig:pareto}
  \end{minipage}

\end{figure*}

We first investigate how accurately \kestrel detects anomalies in both single and mixed anomaly scenarios (Table~\ref{tab:anomalies}).

\indsignpost{Single anomaly scenarios} Figure~\ref{fig:auprc_per_head} compares the area under the precision--recall curve (AUPRC) across the three telemetry approaches for Scenarios 1–4. Detection accuracy varies significantly by anomaly type and telemetry granularity. For \textit{congestion}, all three approaches perform well: \counter achieves $0.88$, while both \dsamp and \kestrel{} reach near-perfect accuracy (AUPRC ${\approx}1.0$). Congestion's clear QFI-level signature makes it detectable even with coarse counters.

\textit{Contention} presents the starkest contrast: \counter collapses to $0.22$, \dsamp improves to $0.37$, while \kestrel{} achieves $0.75$, more than $3\times$ better than \counter and $2\times$ better than \dsamp. This gap stems from contention's anomaly signature: subtle distortions that evade both aggregate counters and threshold-based sampling but are captured by \kestrel{}'s sketches. For \textit{microbursts}, the gap is narrower: \counter reaches $0.48$, while $\Delta$-SMP ($0.86$) and \kestrel{} ($0.87$) perform comparably, both nearly doubling counter accuracy. For \textit{policy abuse}, \dsamp slightly edges out \kestrel{} ($0.93$ vs. $0.90$), although both outperform counters ($0.52$).

\indsignpost{Mixed anomalies} We next consider Scenario~5, where multiple anomaly types may overlap in the same window. This is a strictly harder setting: anomaly signatures dilute one another, and overlap increases the likelihood of false positives. Figure~\ref{fig:auprc_mixed} shows that \counter fails entirely (AUPRC $0.43$, F1 $0.49$), while \dsamp achieves moderate accuracy ($0.71/0.77$). \kestrel{} maintains the highest performance ($0.80/0.81$), which is about $13\%$ higher AUPRC and $5\%$ higher F1 than \dsamp, and nearly $2\times$ better than \counter. \kestrel{}'s advantage stems from continuous summarization of per-packet information, which allow it to capture even very subtle changes in telemetry patterns.

\indsignpost{Take-away} \kestrel{} maintains high accuracy across anomaly types and remains robust to overlapping anomalies, outperforming baselines especially on subtle cases like contention.

\subsection{Cost-Effectiveness} \label{subsec:cost_effectiveness}

We next turn from accuracy to cost, quantifying the telemetry overhead required to achieve the results above. 

\indsignpost{Accuracy-cost trade-off} Figure~\ref{fig:pareto} plots the accuracy-cost trade-off, with ideal operating points in the top-left region. \counter occupies the lower-left: minimal overhead ($\sim$1 Mbps) but poor accuracy (AUPRC $\sim$0.4). \dsamp achieves moderate accuracy ($0.7$--$0.76$) but at very high cost: $60$--$80$ Mbps. \kestrel defines the Pareto frontier, achieving AUPRC $0.7$--$0.82$ with only $5-10$ Mbps export overhead. The sweet spot at $w{=}512,d{=}3$ delivers AUPRC $0.81$ which is $2\times$ higher accuracy than \counter and $10\times$ cheaper than \dsamp. Larger configurations show diminishing returns, confirming our parameter analysis from \S\ref{sec:analysis}.

\indsignpost{Predictable and bounded cost} At the tested loads of 1–4\,Gbps (\S\ref{sec:implementation}), \kestrel{}'s export rate remains stable at ${\sim}$6\,Mbps (0.15–0.6\% of traffic rate).
Overhead depends only on configuration parameters ($w$, $d$, bin count, and number of QIDs) rather than the instantaneous packet rate, making telemetry export cost deterministic and controllable.
In contrast, \dsamp{}'s event-driven postcards scale unpredictably with traffic burstiness and anomaly severity, producing spikes during busy periods.
In our evaluation, \dsamp consumed 60–80\,Mbps (1.5–8\% of traffic rate), an order of magnitude higher despite lower accuracy.

\indsignpost{Take-away} \kestrel{} provides high accuracy with predictable, configuration-bounded overhead, simplifying telemetry planning for operator-scale UPFs.

\subsection{Responsiveness}
\label{subsec:responsiveness}

\begin{figure}[!ht]
\centering
\begin{minipage}[t]{0.48\columnwidth}\vspace{0pt}
\centering
\includegraphics[width=\linewidth]{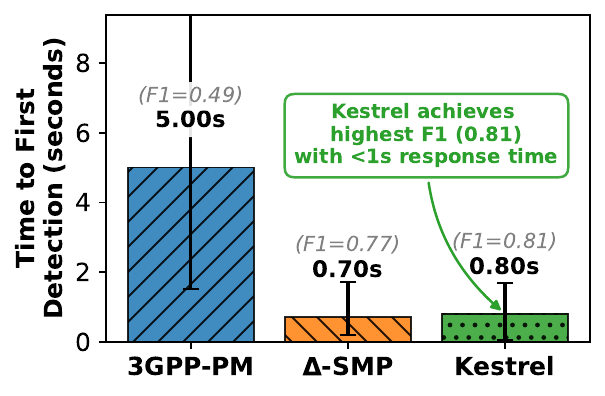}
\end{minipage}
\hfill
\begin{minipage}[t]{0.48\columnwidth}\vspace{0pt}
\centering
\footnotesize
\setlength{\tabcolsep}{4pt}
\begin{tabularx}{\linewidth}{l *{3}{Y}}
\toprule
\textbf{Anomaly} & \textbf{3GPP-PM} & \textbf{$\Delta$-SMP} & \textbf{Kest.} \\
\midrule
Congestion   & 6.00 & 1.20 & 1.05 \\
Contention   & 6.00 & 0.20 & 1.05 \\
Microburst   & 0.50 & 0.20 & 0.05 \\
Policy abuse & 4.00 & 1.20 & 0.55 \\
\bottomrule
\end{tabularx}
\end{minipage}%
\caption{Median time to first detection (left: by scheme with F1 scores; right: by anomaly type).}
\label{fig:ttfd}
\end{figure}

We next evaluate responsiveness, measured as the \emph{time-to-first-detection} (TTFD): the median delay between anomaly onset and the first alarm. Figure~\ref{fig:ttfd} summarizes results across telemetry approaches (left) and per anomaly type (right), using the tuned thresholds from \S\ref{subsec:anomaly_detection}.

\indsignpost{Aggregate performance} Figure~\ref{fig:ttfd} (left) shows that \kestrel{} responds quickly (median $0.80$s) with the highest detection performance (F1 $0.81$). \dsamp responds slightly faster ($0.70$s) but with lower performance (F1 $0.77$). \counter is much slower ($5.00$s) and far less accurate (F1 $0.49$), making it unsuitable for timely anomaly detection.

\indsignpost{Per-anomaly breakdown} For \textit{microbursts}, \kestrel{} reponds immediately ($0.05$s median) while maintaining high F1 ($0.92$). \dsamp also reacts quickly ($0.20$s) but with lower accuracy, missing bursts when metric changes remain below the threshold. \counter responds resonably fast ($0.50$s) but miss smaller events. \dsamp exhibits slightly lower latency in some cases (e.g., $0.20$s for contention vs. $1.05$s for \kestrel{}). This is because postcards are exported immediately upon threshold crossings, while \kestrel{} waits for the end of each window to query sketches. However, this speed advantage comes at the cost of detection fidelity: for \textit{contention}, \dsamp's AUPRC is only 0.37 (recall \S\ref{subsec:anomaly_detection}) compared to \kestrel{}'s $0.75$, demonstrating that faster alarms are only valuable when reliable. For \textit{congestion}, both achieve similar latency ($1.05$–$1.20$s) with high accuracy. For \textit{policy abuse}, sketches detect anomalies within $0.55$s, roughly ${7\times}$ faster than \counter ($4.00$s) and ${2\times}$ faster than \dsamp ($1.20$s).

\indsignpost{Take-away} \kestrel{} balances responsiveness and accuracy: while per-packet export can react instantly to anomaly events, \kestrel{}'s windowed summarization achieves comparable latency with higher detection fidelity.


\subsection{Microbenchmarks and Design Insights} \label{subsec:microbenchmarks}

\begin{figure*}[!ht]
\centering

\begin{subfigure}[b]{0.32\textwidth}
\centering
\small
\begin{tabular}{lccc}
\toprule
\textbf{Scheme} & \textbf{p50} & \textbf{IQR} & \textbf{I/F/ML} \\
& \textbf{(ms)} & \textbf{(± ms)} & \textbf{(\%)} \\
\midrule
3GPP-PM   & 1.83  & 0.04 & 37/29/33 \\
Kestrel   & 53.3  & 0.45 & 27/72/1 \\
$\Delta$-SMP & 200.4 & 7.3 & 33/66/1 \\
\bottomrule
\end{tabular}
\subcaption{Per-window pipeline latency}
\label{fig:pipeline_latency}
\end{subfigure}
\hfill
\begin{subfigure}[b]{0.32\textwidth}
\centering
\vspace{0pt}
\includegraphics[width=\linewidth]{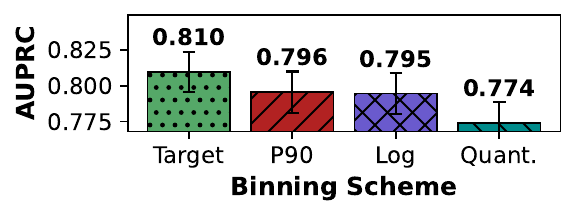}
\subcaption{Accuracy across binning strategies}
\label{fig:auprc_bins}
\end{subfigure}
\hfill
\begin{subfigure}[b]{0.32\textwidth}
\centering
\small
\begin{tabular}{lr}
\toprule
\textbf{Resource} & \textbf{Usage} \\
\midrule
Memory (SRAM/TCAM) & $<$20\% \\
Hashing (Hash Units) & $\approx$25\% \\
Computation (ALU) & $\approx$35\% \\
\bottomrule
\end{tabular}
\subcaption{Switch resource footprint}
\label{fig:kestrel_resources}
\end{subfigure}

\caption{Microbenchmarks of \kestrel{}. (a) Per-window pipeline latency, showing bounded cost. (b) Accuracy across binning strategies, target-occupancy performs best. (c) Switch resource footprint, showing modest usage.}
\label{fig:impl_results}
\end{figure*}

We next present microbenchmarks validating \kestrel{}'s practicality, covering (a) per-window processing latency, (b) binning sensitivity, and (c) Tofino resource usage (Fig.~\ref{fig:impl_results}).

\indsignpost{Pipeline latency} 
We measure end-to-end processing per $1$s window, broken down into ingestion (I), feature extraction (F), and ML detection (ML). Figure~\ref{fig:impl_results}(a) shows that all methods finish well within the deadline: counters in $\sim$2\,ms, \kestrel{} in $53$ms, and \dsamp in $200$ms (median).  Counters are trivially cheap ($\sim2$ms) but have poor performance. 
\kestrel spends most of its budget ($72\%$) on feature extraction, as sketch registers must be queried and aggregated, yet the absolute cost remains modest: $\sim53$ms per window, independent of packet rate or number of flows. 
In contrast, \dsamp spends two-thirds of its budget on feature extraction because each sampled packet is processed individually, causing latency to grow with traffic volume. \kestrel also exhibits low variance (IQR $\pm0.45$ms), confirming predictable performance. Therefore, it strikes a practical middle ground: richer than counters, yet an order of magnitude faster and more stable than postcards.

\indsignpost{Binning strategies} A key design choice in \kestrel{} is how to discretize latency and IAT distributions into bins. We evaluate four approaches that represent different philosophies for capturing distributional behavior: (1) \textit{Target}, \kestrel{}'s target-occupancy method (\S\ref{subsec:binning}), which caps baseline occupancy in diagnostic bins at~$\rho\%$ to make anomalies appear as sharp spikes;  (2) \textit{P90}, which sets boundaries at the 90th and 99.8th percentiles with log spacing to emphasize the ``knee'' of latency/IAT distributions; (3) \textit{Log}, which applies uniformly log-spaced bins across a fixed global range, capturing both short and long tails without per-QID tuning; and   (4) \textit{Quantile}, which divides samples into equal-frequency bins for uniform sensitivity across the range.  
Target-occupancy achieves the highest AUPRC ($0.810$) compared to P90 ($0.796$), Log ($0.795$), and Quantile ($0.774$). Although these improvements seem modest, they have a practical impact on anomaly detection: target-occupancy reduces contention false positives by ${\sim}20\%$ at $90\%$ recall, compared to other binning approaches.

\indsignpost{Hardware usage} Figure~\ref{fig:impl_results}(c) shows \kestrel{}'s resource footprint on Tofino. The design consumes less than 20\% of SRAM/TCAM, about 25\% of hash units, and 35\% of ALUs, leaving substantial headroom for other UPF functions. 
These results demonstrate that Kestrel fits comfortably within the constraints of a commercial programmable switch.

\indsignpost{Take-away} \kestrel{}'s principled, distribution-aware design translates into practical gains: bounded latency, efficient resource use, and reliable anomaly detection on real hardware.

\section{Related Work} \label{sec:related_work}

\signpost{Telemetry paradigms} \emph{In-band telemetry (INT)} encodes metadata directly in packets~\cite{int-spec-v2.1,hpcc-sigcomm19,pint2020probabilistic,tang2022orchestration,sheng2021deltaint}. While widely used in datacenters, traditional INT is challenging to deploy in mobile networks due to GTP-U encapsulation, middlebox handling, and lack of end-to-end programmability.  
\emph{Event-driven approaches} report telemetry only when certain predicates are met. For example, Marple~\cite{marple-sigcomm17}, BurstRadar~\cite{burstradar-apsys18}, and NetSeer~\cite{netseer-sigcomm20} trigger telemetry on queue thresholds or on drops, reducing cost but limiting visibility to predefined events.  \emph{Sketch-based approaches} maintain compact summaries~\cite{opensketch,elasticsketch,nitrosketch}, and have been previously applied to scenarios such as heavy hitter and volumetric attack detection. However, prior works, typically do not focus on \emph{per-flow distributions} of latency, IAT, or policing colors, features that are critical for QoS anomaly detection at the UPF. \kestrel{} addresses precisely this gap.

\indsignpost{Anomaly detection} Prior anomaly detection efforts focus on control-plane behavior and aggregate metrics. For the 5G core, approaches include deep sequence models to identify abnormal network function (NF) interactions \cite{anomaly-detection-tccn25}, AI/ML for detecting anomalous traffic events \cite{nwdaf-globecom23}, and detection of service degradation in cloud-based NF deployments \cite{anomaly-detection-tnsm23}. In the RAN context, prior works have investigated distributed anomaly detection models across edge and cloud~\cite{spotlight-mobicom24}, as well as signaling storms and their mitigation~\cite{signalling-storm-commag24}. Slice-level anomaly detection \cite{ddos-globecom22,anomaly-detection-noms23} addresses security and QoS threats at the granularity of network slices. For example, \cite{ddos-globecom22} applies deep learning to detect DDoS attacks targeting slices and to dynamically isolate attackers,while \cite{anomaly-detection-noms23} proposes a graph-based framework that correlates multivariate slice KPIs to proactively detect and explain anomalies. These approaches operate either on packet captures, incurring very high cost, or on aggregate counters and KPIs which cannot capture transient per-TEID
QoS violations within a slice. \kestrel{} complements these efforts by offering fine-grained, data plane visibility, enabling the detection of QoS anomalies that precede or evade slice-level KPI deviations.

\section{Conclusion}

We presented \kestrel{}, a principled telemetry system for detecting QoS anomalies in 5G user planes. 
By capturing distributional signals such as latency tails and inter-arrival patterns within compact, anomaly-aware sketches, 
\kestrel{} provides fine-grained visibility at minimal export cost. 
Our Tofino testbed evaluation shows that it delivers high detection accuracy, sub-second responsiveness, and predictable hardware utilization, 
demonstrating that anomaly-driven telemetry design is both practical and effective for next-generation mobile networks.

\itsignpost{Future work} While this work focused on hardware-accelerated UPFs, the same principles of anomaly-driven telemetry could extend to software data planes (e.g., eBPF or VPP), enabling lightweight, anomaly-aware monitoring for heterogeneous 5G deployments.

\bibliographystyle{IEEEtran}
\bibliography{references}

\appendix
\section*{Appendix A \; Proofs}

This appendix provides proofs for the formal guarantees and sizing rules in \S\ref{sec:analysis}.

\indsignpost{Proof of Detectability Condition (with spillover)}
By the CMS guarantee, each bin estimate satisfies  
$x_{k,j} \le \hat{x}_{k,j} \le x_{k,j} + \varepsilon N_j$ with probability at least $1-\delta$.  
Summing over the diagnostic bins $T$ gives:
\[
x_{k,T} \le \hat{x}_{k,T} \le x_{k,T} + \varepsilon N_T,
\]
and summing over all bins gives:
\[
x_k \le \hat{x}_k \le x_k + \varepsilon N'.
\]
In baseline conditions, the maximum estimated diagnostic ratio is
\[
\frac{\hat{x}_{k,T}}{\hat{x}_k} \le \frac{x_{k,T} + \varepsilon N_T}{x_k}.
\]
During an anomaly adding $\Delta_T$ packets to $T$ and $\beta\Delta_T$ elsewhere, 
the diagnostic mass increases while the total mass grows modestly.  
The minimum estimated ratio becomes
\[
\frac{\hat{x}_{k,T}}{\hat{x}_k} \ge \frac{x_{k,T} + \Delta_T}{x_k + \Delta_T + \beta\Delta_T + \varepsilon N'}.
\]
Detection succeeds when the anomaly-window ratio exceeds the baseline maximum.  
Rearranging this inequality yields the detectability condition.
\qed

\indsignpost{Proof of Width Requirement}
In the sparse diagnostic regime ($x_{k,T}\ll x_k$, $N_T\ll N'$), the detectability condition simplifies to
\[
(1-\beta)\Delta_T \gtrsim \varepsilon N_T, \quad \text{where } \varepsilon \approx e/w.
\]
Enforcing the design constraints $N_T \le N_T^{\max}$, $\Delta_T \ge \Delta_T^{\min}$, 
and $\beta \le \beta_{\max}$ gives
\[
(1-\beta_{\max})\Delta_T^{\min} \ge \frac{e}{w} N_T^{\max}.
\]
Solving for $w$ yields the width requirement.
\qed

\indsignpost{Proof of Depth Requirement}
Each CMS estimate violates its error bound with probability $\le e^{-d}$.  
Detection involves $K$ diagnostic-bin estimates and one total-count estimate ($K{+}1$ events total).  
By union bound, all bounds hold simultaneously with probability $\ge 1-(K+1)e^{-d}$.  
Setting this $\ge 1-\zeta$ gives
\[
d \ge \left\lceil \ln\!\left(\frac{K+1}{\zeta}\right) \right\rceil.
\]
\qed

\indsignpost{Proof of Target-Occupancy Proposition}
Let $F$ be the baseline CDF.  
Placing the latency-tail boundary at the $(1-\rho)$-quantile ensures exactly $\rho N$ packets have latency above that boundary.  
Similarly, placing the IAT-head boundary at the $\rho$-quantile ensures $\rho N$ packets have IAT below it.  
Thus, $N_T^{\max} = \rho N$.
\qed

\indsignpost{Proof of Drift Effect Proposition}
From the width requirement, required width scales as $w \propto N_T^{\max}$.  
If diagnostic occupancy increases from $\rho N$ to $\rho'N$, then 
$w_{\text{new}}/w_{\text{old}} = \rho'/\rho$.
\qed

\indsignpost{Practical Implications}
These results explain the parameter choices in \S\ref{sec:sketch_parameters}: 
width $w$ scales linearly with $N_T^{\max}$ but inversely with $\Delta_T^{\min}$, 
while depth $d$ grows only logarithmically with the number of diagnostic bins~$K$.  
This justifies why moderate parameters ($w{=}512$, $d{=}3$) suffice in practice, 
as empirically validated in \S\ref{subsec:cost_effectiveness}.

\end{document}